# Brain-Computer Interface with Corrupted EEG Data: A Tensor Completion Approach

J. Solé-Casals [1,*], C. F. Caiafa [2,3], Q. Zhao [4,5], A. Cichocki [6,7,8]

[1] University of Vic – Central University of Catalonia, Data and Signal Processing Research Group, Vic, Catalonia (Spain).
[2] Instituto Argentino de Radioastronomía (IAR) – CCT-La Plata, CONICET; CICPBA, Villa Elisa, Buenos Aires, Argentina.
[3] Department of Psychological and Brain Sciences – Indiana University, Bloomington, Indiana, United States.
[4] Tensor Learning Unit – RIKEN Center for Advanced Intelligence, Tokyo, Japan.
[5] School of Automation, Guangdong University of Technology, China.
[6] Skolkovo Institute of Science and Technology, Moscow, Russia.
[7] Department of Informatics, Nicolaus Copernicus University, 87-100 Torun, Poland.
[8] College of Computer Science, Hangzhou Dianzi University, 310018 Hangzhou, China.

[*] Corresponding author: jordi.sole@uvic.cat

**Abstract**

*Background / Introduction:* One of the current issues in Brain-Computer Interface (BCI) is how to deal with noisy Electroencephalography (EEG) measurements organized as multidimensional datasets (tensors). On the other hand, recently, significant advances have been made in multidimensional signal completion algorithms that exploit tensor decomposition models to capture the intricate relationship among entries in a multidimensional signal. We propose to use tensor completion applied to EEG data for improving the classification performance in a motor imagery BCI system with corrupted measurements. Noisy measurements (electrode misconnections, subject movements, etc.) are considered as unknowns (missing samples) that are inferred from a tensor decomposition model (tensor completion).

*Methods:* We evaluate the performance of four recently proposed tensor completion algorithms, CP-WOPT (Acar et al, 2011), 3DPB-TC (Caiafa et al, 2013), BCPF (Zhao et al, 2015) and HaLRT (Liu et al. (2013), plus a simple interpolation strategy, first with random missing entries and then with missing samples constrained to have a specific structure (random missing channels), which is a more realistic assumption in BCI applications.

*Results:* We measured the ability of these algorithms to reconstruct the tensor from observed data. Then, we tested the classification accuracy of imagined movement in a BCI experiment with missing samples. We show that for random missing entries, all tensor completion algorithms can recover missing samples increasing the classification performance compared to a simple interpolation approach. For the random missing channels case, we show that tensor completion algorithms help to reconstruct missing channels, significantly improving the accuracy in the classification of motor imagery (MI), however, not at the same level as clean data. Summarizing, compared to the interpolation case, all tensor completion algorithms succeed to increase the classification performance by 7 – 9% (LDA – SVD) for random missing entries and 15 - 8% (LDA – SVD) for random missing channels.

*Conclusions:* Tensor completion algorithms are useful in real BCI applications. The proposed strategy could allow using motor imagery BCI systems even when EEG data is highly affected by missing channels and/or samples, avoiding the need of new acquisitions in the calibration stage.

## 1. Introduction

Neural signal processing combines machine learning and computational methods with the aim of improving BCI in real-like scenarios. BCI, acting as an interface, could enable normal function in cases of brain disease or injury, and provide brain monitoring, or medical rehabilitation of brain disorders [1]. Therefore, BCI applications are becoming more and more popular in recent years [2]. As defined in [3], "A BCI is a communication system that does not depend on the brain's normal output pathways of peripheral nerves and muscles". In other words, a BCI is a system that allows a person to communicate only using his/her brainwaves, without any peripheral muscular activity [4]. For example, BCI systems can be used for controlling a wheelchair [5], [6], or a prosthetic arm [7], for computer access [8], or simply for communication for people affected by a number of motor disabilities [9], [10]. Virtually anything that can be controlled by a computer could, potentially, be controlled by a BCI [11], [12].

To use a BCI, the subject must generate specific brain activity patterns that can be detected by machine learning methods, which can translate neuroimaging patterns to specific commands. The mechanisms by which the specific brain activity patterns are generated depend on BCI paradigms. There exist several paradigms for BCI that can be distinguished based on the used neuroimaging methods and control signals [13]. Standard control signals include: Visual Evoked Potentials (VEP), Slow Cortical Potentials (SCP), P300 Evoked Potentials and Motor Imagery (see **Table 1**) [14]. In this work, we focus only on Electroencephalography (EEG) based methods and the Motor Imagery paradigm.

In the Motor Imagery (MI) paradigm, the subject imagines the movement of one of his/her limbs (hands in our experiments) for a few seconds. This imagination activates brain areas in the motor cortex, which are similar to those that are activated when the actual movement of the same limb is executed. In our experiments, a BCI application based on the motor imagery paradigm allows us to choose between two actions depending on whether the subject is imagining the movement of his/her left or right hand.

*<<TABLE 1 HERE>>*

*1.1. The problem of data corruption in EEG recordings*

When recording EEG signals, corrupted data can be originated from a high impedance value between electrodes and the scalp, or even worse, from one electrode being detached during the recording session. Also, some body movements of the subject can cause huge artefacts in the EEG signal [15], [16]. For example, eye blinks or jaw movements can cause contamination of EEG signals of high amplitude affecting all the electrodes [17], [18]. Other less common problems can be generated by improper electrical connections between electrodes and amplifier, malfunction of the amplifier or the A/D converter, environmental noise, etc.

Data corruption is an important problem to solve, for example, in the calibration step of MI BCIs where classifiers need as many trials as possible to learn the unique features of each new user's EEG activity. Typically, one can decide to eliminate the corrupted channel(s), or the whole experiment (trial). However, EEG data is precious and discarding it is not an option in some scenarios. For example, if we are dealing with a BCI system based in sensorimotor rhythms, online discarding a trial means neither feedback nor command output, which could confuse the user and generate frustration. When data corruption is restricted to isolated samples, the straightforward method to solve is to infer corrupted/missed samples using simple interpolation or regression approach. When a complete channel measurement is missing, one may use the measurements taken in different trials for that channel (or neighbour channels) to reconstruct it by simple interpolation. In this work we propose to use of tensor completion algorithms because they are able to capture the multidimensional structure of measurements and the complex relationship between channels, time and trials, which is crucial to obtain optimal reconstruction of missing or corrupted samples.

*1.2. Tensor decomposition notation and definitions*

Tensors generalize vectors (1D arrays, $\mathbf{y} \in \mathbb{R}^I$) and matrices (2D arrays, $\mathbf{Y} \in \mathbb{R}^{I_1 \times I_2}$) to arrays of higher dimensions (ND arrays, $\boldsymbol{\mathcal{Y}} \in \mathbb{R}^{I_1 \times I_2 \times \cdots \times I_N}$), three or more ($N \geq 3$). Such arrays can be used to perform multidimensional factor analysis and decomposition and are of interest to many scientific disciplines including signal processing and machine learning [19]. The success of tensor decomposition models is based in their ability to capture a reduced number of latent structures that explain multidimensional datasets.

Entries $(i, j, k)$ of a tensor $\boldsymbol{\mathcal{Y}} \in \mathbb{R}^{I \times J \times K}$ are referred to as $\boldsymbol{\mathcal{Y}}(i,j,k)$ or $y_{ijk}$. Given a matrix $\mathbf{A} \in \mathbb{R}^{I \times R}$, the vectors $\mathbf{a}_i \in \mathbb{R}^R$ and $\mathbf{a}^j \in \mathbb{R}^I$ are obtained by extracting the $i$-row and the $j$-column from matrix $\mathbf{A}$.

Below, we introduce the tensor decomposition models used in this work:

*1.2.1. CANDECOMP/PARAFAC (CP):* In this model, a data tensor $\boldsymbol{\mathcal{Y}} \in \mathbb{R}^{I \times J \times K}$ is approximated by a sum of rank-1 tensors, where the rank $R$ is given by the number of terms (**Figure 1(a)**):

$$\boldsymbol{\mathcal{Y}}(i,j,k) \approx \sum_{r=1}^{R} a_{ir} b_{jr} c_{kr}, \tag{1}$$

1.2.2. *Sparse Tucker decomposition:* This is an extension of the classical Tucker model and [20] has been introduced in [20, 21]. In this model, a data tensor $\boldsymbol{\mathcal{Y}} \in \mathbb{R}^{I \times J \times K}$ is approximated by multiplying a "large" sparse core tensor $\mathcal{X} \in \mathbb{R}^{R_1 \times R_2 \times R_3}$ ($R_1 \geq I$, $R_2 \geq J$, $R_3 \geq K$) by factor matrices in each dimension: $\boldsymbol{D}_1 \in \mathbb{R}^{I \times R_1}$, $\boldsymbol{D}_2 \in \mathbb{R}^{J \times R_2}$ and $\boldsymbol{D}_3 \in \mathbb{R}^{K \times R_3}$, as follows:

$$\mathcal{Y}(i,j,k) \approx \sum_{r_1=1}^{R_1} \sum_{r_2=1}^{R_2} \sum_{r_3=1}^{R_3} \mathcal{X}(r_1, r_2, r_3) \, \boldsymbol{D}_1(i, r_1) \boldsymbol{D}_2(j, r_2) \boldsymbol{D}_3(k, r_3), \qquad (2)$$

which can be written in the following compact form (**Figure 1(b)**):

$$\mathcal{Y} \approx \mathcal{X} \times_1 \boldsymbol{D}_1 \times_2 \boldsymbol{D}_2 \times_3 \boldsymbol{D}_3. \qquad (3)$$

Factor matrices $\boldsymbol{D}_1$, $\boldsymbol{D}_2$ and $\boldsymbol{D}_3$ are called dictionaries of atoms because the original signal $\mathcal{Y}$ is represented as a superposition of a reduced number of rank-1 tensors. Moreover, by vectorising equation (**14**), we obtain the following expression:

$$\boldsymbol{y} \approx \boldsymbol{D}\boldsymbol{x}, \ \text{ with } \boldsymbol{D} = \boldsymbol{D}_3 \otimes \boldsymbol{D}_2 \otimes \boldsymbol{D}_1, \qquad (4)$$

where $\otimes$ stands for the Kronecker product and $\boldsymbol{D} \in \mathbb{R}^{IJK \times R_1 R_2 R_3}$ is referred as a global dictionary [20].

<<FIGURE 1 HERE>>

### *1.3. Inference of missing entries in tensor data*

The problem of estimating missing data can arise in a variety of real-world applications with multidimensional (tensor) structured data, which is referred as tensor completion. In this work, we propose to use tensor completion approach to corrupted EEG data. For this purpose, we organize EEG data as a 3D tensor whose dimensions are channel, time and trial ($channel \ x \ time \ x \ trial$) and consider every corrupted sample as a missing entry. Thus, we estimate missing entries by applying a tensor completion algorithm.

Several tensor completion strategies have been developed recently for different applications. For example, an underlying tensor factorization model [19] with a fixed rank is assumed and the objective is to infer the underlying factors from partially observed data. In [22], a CANDECOMP/PARAFAC (CP) factorization with missing data was formulated as a weighted least squares problem, termed CP weighted optimization (CP-WOPT). Some other related methods were also investigated such as the structured CP decomposition using nonlinear least squares (CPNLS) [23]. However, these tensor factorization schemes can be prone to overfitting due to an imprecise tensor rank selection and a point estimation of latent factors rather than learning posterior distributions, resulting in severe deterioration of predictive performance [24]. Recently, the Bayesian CP factorization was proposed by imposing column-wise sparsity over factor matrices, resulting in an automatic tensor rank determination by model inference. In addition, the fully Bayesian approach can effectively prevent overfitting problem and avoid manually tuning parameters [24].

Other tensor completion techniques are inspired by matrix inpainting algorithms by exploiting an automatic rank optimization as a convex optimization on a tensor nuclear norm [25]. Some variants were also proposed under this framework, such as fast composite splitting algorithms (FCSA) [26]. Since these completion-based methods cannot explicitly capture the underlying factors, a simultaneous tensor decomposition and completion (STDC) [27] method was proposed in which multilinear rank minimization was applied for Tucker decomposition. It is also noteworthy that the rank minimization based on nuclear norm is sensitive to tuning parameters, which may tend to over- or under-estimate the true tensor rank.

Finally, another option for tensor completion is to assume that every small patch of the tensor admits a sparse decomposition over a known dictionary such as a Kronecker Wavelet or the Discrete Cosine Transform (DCT) bases. Moreover, it is known that better results can be obtained by training a dictionary based on available datasets. A patch-based tensor completion technique with a Kronecker structured dictionary learning algorithm has been recently proposed in [21].

Is in this paper, we evaluate four tensor completion techniques applied to multichannel EEG data, allowing us to keep all the recorded information in a tensor format (multidimensional) and to recover the corrupted entries by means of tensor completion algorithms. The choice of the algorithms was based on the fact that they represent four different approaches to the problem of tensor completion, from simple least squares optimization (CP-OPT), through more sophisticated algorithms based on local sparse representations (3DPB-TC), low-rank Bayesian approach (BCPF) and trace-norm minimization strategy (HaLTRC).

## 2. Methods

### 2.1. Experimental paradigm

The EEG dataset used in this work was provided by the Laboratory for Advanced Brain Signal Processing, BSI-RIKEN, Japan, in collaboration with Shanghai Jiao Tong University, China. The dataset is freely available at: http://mon.uvic.cat/data-signal-processing/en/descarregues/. This dataset was recorded for 5 healthy subjects. The cue-based BCI paradigm consisted of two motor imagery tasks, namely the imagination of movement of the left hand (LH) and right hand (RH). Several sessions on different days were recorded for some subjects (**Table 2**).

<<FIGURE 2 HERE>>

<<TABLE 2 HERE>>

The experiment is illustrated in **Figure 2(a)**. The subjects were sitting in a comfortable armchair in front of a computer screen. At the beginning of a trial, the screen is blank. After two seconds ($t = 2s$), a cue in the form of an arrow pointing either to the left or right (corresponding to two classes of LH and RH) appeared and remained on the screen for a specific duration (3-4sec). This prompted the subjects to perform the desired motor imagery task. The subjects were requested to carry out the motor imagery task until the cue disappeared from the screen and try to avoid eye blinking or eye movements during the imagination. When the cue disappears, a two seconds' break follows. This procedure is repeated 80 -170 times for each session with the random cue sequence.

For each subject, the first run is called initialization procedure, which only presents the cues without any feedback. Based on the online BCI classifier trained on the EEG data recorded from the initialization run, the system can give a feedback online by showing several red bars representing the classification output for left and right hand commands. Meanwhile, the EEG data with class labels are recorded.

*2.2. Data recording and format*

In this dataset, g.tec (g.USBamp) was used for recording the EEG signals. The EEG signals were band-pass filtered between 2Hz and 30Hz with sample rate of 256Hz and a notch filter at 50Hz was applied.

The signals are measured in µV. The number of electrodes was fixed to 6 in all the experiments in order to develop a BCI system with number of electrodes as small as possible. The selected electrodes are C3, Cz, C4, CP3, CPz and CP4, and its arrangement is shown in **Figure 2(b)**, highlighted in blue

The experiments conducted in different days for the same subject are called different sessions. Each file contains one session, which is comprised of several trials separated by short breaks. Subject C has many sessions conducted on different days, while for the rest of them we only have data from one session. The EEG data stored in each data set is organized in segment structure; each segment represents a single trial with one specific class label. The EEG data is provided as a 3D array (tensor) with size of [$N_{channel}$ x $N_{time}$ x $N_{trial}$]. '$N_{channel}$' denotes the number of electrodes, '$N_{time}$' is the number of time samples in each imagination task and '$N_{trial}$' is the number of motor imagery tasks performed in this session. **Table 2** summarizes the information of all used datasets.

*2.3. Tensor Completion Algorithms*

The problem of estimating missing entries in higher dimensional arrays has been widely studied in recent years and there is a long list of available algorithms to choose from [22, 20, 27, 28, 25, 29, 24, 30, 31] [32, 33, 34, 35, 36]. Tensor completion has been recently proposed for handling missing measurements before classification, for example, in human activity recognition [37, 38]. In this work, we propose to use it for handling missing samples during calibration of a BCI system. Four different tensor completion algorithms were tested in the experiments (see summary in **Table 3**). The choice of the algorithms was based on the fact that they represent four different approaches to the tensor completion problem, from a simple least squares optimization (*CP-OPT*), through more sophisticated algorithms based on local sparse representations (*3DPB-TC*), low-rank Bayesian approach (*BCPF*) and trace-norm minimization strategy (*HaLTRC*). We also used a basic interpolation algorithm (*AIaT*) to estimate missing entries and used as a reference to evaluate the quality of the results given by all the algorithms. A brief description of each algorithm is detailed below.

<<TABLE 3 HERE>>

*2.3.1. CANDECOMP/PARAFAC Weighted Optimization Algorithm (CP-WOPT)*

CANDECOMP/PARAFAC (CP) algorithms are the simplest and most well know tensor factorization that captures multi-linear structure [19]. In the presence of missing data, CP can be formulated as a weighted least squares problem that models only the known entries. CP-WOPT [22] is an algorithm that uses first-order optimization to solve the weighted least squares objective function over all the factor matrices simultaneously. As detailed in [22], the method uses a weighted version of the error function to ignore missing data and model only the known entries. In that case, nonlinear optimization can be used to directly solve the weighted least squares problem for the CP model. The weighted version of the error function to be minimized is

$$f_W(A, B, C) = \frac{1}{2} \sum_{i=1}^{I} \sum_{j=1}^{J} \sum_{k=1}^{K} \left\{ w_{ijk} \left( y_{ijk} - \sum_{r=1}^{R} a_{ir} b_{jr} c_{kr} \right) \right\}^2 \qquad (5)$$

where $\mathcal{Y}$ is a three-way tensor of size $I \times J \times K$ and rank $R$; $A$, $B$ and $C$ of sizes $I \times R$, $J \times R$ and $K \times R$, respectively, are the factors matrices in the CP model such that

$$y_{ijk} \approx \sum_{r=1}^{R} a_{ir} b_{jr} c_{kr}, \quad \text{for all } i = 1, \dots, I, j = 1, \dots, J, k = 1, \dots, K$$

and tensor $\mathcal{W}$, which is the same size as $\mathcal{Y}$, with entries defined as

$$w_{ijk} = \begin{cases} 1 & \text{if } y_{ijk} \text{ is known} \\ 0 & \text{if } y_{ijk} \text{ is missing} \end{cases} \quad \text{for all } i = 1, \dots, I, j = 1, \dots, J, k = 1, \dots, K.$$

See [22] for a detailed explanation of the CP-WOPT algorithm.

*2.3.2. 3D Patch Based Tensor Completion Algorithm (3DPB-TC)*

The 3D Patch-Based Tensor Completion method consists in dividing the whole, and potentially large, 3D tensor $\mathcal{Y}$ in a large collection of small overlapped 3D patches (subtensors): $\mathcal{Y}_1, \mathcal{Y}_2, \dots, \mathcal{Y}_N$ In this method, it is assumed that the vectorized version $y$ of every 3D patch has a sparse representation on a known dictionary matrix (see equation (**4**)), i.e. $y \approx Dx$, where $y$ is the vector obtained by concatenating all entries of the 3D-patch in a long vector, $D$ is a matrix containing in its columns "atoms" or prototype signals, and $x$ is a sparse vector whose non-zero entries indicate which "atoms" are linearly combined to obtain an approximation of $y$.

It is known that signals having a sparse representation on a known dictionary can be recovered from a small subset of its entries, which is one of the assumptions in the Compressed Sensing theory [39]. Sparse representations have demonstrated to be useful for inferring missing samples in 2D-images and many algorithms were recently developed based on these ideas [40, 29, 28] but these methods become computationally too expensive when they are extended to 3D-signals (tensors). To overcome this problem, in [21], a method based on a sparse Tucker decomposition model, here referred as the 3DPB-TC algorithm, was proposed. In this algorithm, small dictionaries associated to each one of the signal dimension or mode are used instead of having a large global dictionary. In this work, an alternate least squares algorithm was proposed for learning an optimal set of dictionaries from an available collection of 3D-patches. We refer the reader to [21] work for more details of this algorithm.

*2.3.3. Bayesian CP Factorization (BCPF)*

The BCPF [24] is a generative probabilistic model for CP tensor factorization of incomplete tensor data, which can determine the tensor rank automatically by employing a hierarchical sparsity-inducing prior over latent matrices and a fully Bayesian inference framework. In addition, the predictive uncertainty can be also inferred by BCPF. The observation model of BCPF, i.e. the probability of observed data given the CP factors and parameter $\tau$, is described by

$$p(\mathcal{Y}_\Omega|\{A,B,C\},\tau) = \prod_{i=1}^{I}\prod_{j=1}^{J}\prod_{k=1}^{K} \mathcal{N}\big(y_{ijk}|\langle a_i,b_j,c_k\rangle,\tau^{-1}\big)^{w_{i_1 i_2 i_3}} \qquad (6)$$

where $\mathcal{Y}$ is an incomplete 3th-order tensor of size $I \times J \times K$, $\mathcal{W}$ is a binary tensor of the same size of $\mathcal{Y}$ which indicates the observed entries (the mask, as defined in previous section), $\Omega$ denotes a set of 3-tuple indices $(i,j,k)$, $\{A,B,C\}$ are the factor matrices, $\mathcal{N}(y|\mu,\tau^{-1})$ is a Gaussian distribution with mean $\mu$ and standard deviation $\tau$, and $\langle a,b,c\rangle$ denotes a generalized inner product of three vectors. The sparsity inducing prior for factor matrix $A$ (and equivalently for the other factors $B$ and $C$) is given by

$$p(A|\lambda) = \prod_{i=1}^{I} \mathcal{N}(a_i|0,\Lambda^{-1}) \qquad (7)$$

$$p(\lambda) = \prod_{r=1}^{R} \text{Ga}(\lambda_r|c_0^r, d_0^r) \qquad (8)$$

where $\text{Ga}(.|.)$ denotes a Gamma distribution, which results in a column-wise sparsity property. Additionally, the prior for the noise parameter is given by

$$p(\tau) = \text{Ga}(\tau|a_0, b_0) \qquad (9)$$

By applying the variational Bayesian inference, BCPF allows us to obtain the posterior distribution of all unknown model parameters $\Theta$ as presented by

$$p(\Theta|\mathcal{Y}_\Omega) = \frac{p(\Theta,\mathcal{Y}_\Omega)}{\int p(\Theta,\mathcal{Y}_\Omega)\,d\Theta} \qquad (10)$$

and the predictive distribution over missing data as presented by

$$p(\mathcal{Y}_{\backslash\Omega}|\mathcal{Y}_\Omega) = \int p(\mathcal{Y}_{\backslash\Omega}|\Theta)p(\Theta|\mathcal{Y}_\Omega)\,d\Theta \qquad (11)$$

where, for simplicity of notation, all unknowns including latent variables and hyperparameters are collected and denoted together by $\Theta = \{A,B,C,\lambda,\tau\}$, and the predictive distribution over the missing entries is denoted by $\mathcal{Y}_{\backslash\Omega}$.

BCPF method is characterized as a tuning parameter-free approach and all model parameters can be inferred from the observed data, which avoids the computational expensive parameter selection procedure. In contrast, the other existing tensor factorization methods require a predefined rank, while the tensor completion methods based on nuclear norm require to select several tuning parameters. The initialization point of tensor rank is usually set to its maximum possible value, while the effective dimensionality can be inferred automatically under a Bayesian inference framework. We refer the reader to [24] work for more details of this algorithm.

*2.3.4.* High accuracy Low Rank Tensor Completion (*HaLRTC*)

The HaLRTC algorithm was introduced by Liu et al in [25] together with other two algorithms: Simple Low Rank Tensor Completion (SiLRTC) and Fast Low Rank Tensor Completion (FaLRTC) algorithms. HaLRTC has demonstrated to attain the performance when it was applied to visual data [25]. It generalized the low-rank matrix completion problem [41] to higher dimensions based on a generalized trace norm. The trace norm of a matrix $X \in \mathbb{R}^{I \times J}$ is defined as the sum of its singular values, i.e. $\|X\|_* = \sum_{i=1}^{\min(I,J)} \sigma_i$. The generalized trace norm of a tensor $\mathcal{X} \in \mathbb{R}^{I \times J \times K}$ is defined as a convex combination of the trace norms of the corresponding mode-*n* unfolding matrices:

$$\|\mathcal{X}\|_* = \sum_{n=1}^{3} \alpha_n \|X_{(n)}\|_*, \quad (12)$$

where $X_{(n)}$ is the mode-*n* unfolding matrix obtained from tensor $\mathcal{X}$, i.e. $X_{(1)} \in \mathbb{R}^{I \times JK}$, $X_{(2)} \in \mathbb{R}^{J \times IK}$ and $X_{(3)} \in \mathbb{R}^{K \times IJ}$, and $\alpha_n \geq 0$ s.t. $\sum_{n=1}^{3} \alpha_n = 1$.

The HaLRTC algorithm aims to solve the following optimization problem:

$$\min_{\mathcal{Y}} \|\mathcal{Y}\|_*, \quad \text{s.t. } \mathcal{Y}_\Omega = \mathcal{T}_\Omega, \quad (13)$$

where $\mathcal{Y}_\Omega$ is the set of entries of $\mathcal{Y}$ restricted to the available measurements $\mathcal{T}_\Omega$. The HaLRTC algorithm applies the Alternating Direction Method of Multipliers (ADMM) algorithm [42] to solve the above problem.

*2.3.5.* Average Interpolation across Trials *(AIaT)*

The simplest way to recover random missing entries and random missing channels is calculating the mean using the information contained in other trials. Therefore, we also used the average interpolation across trials as the fifth algorithm, to be compared with the four tensor completion algorithms. The random missing entries were estimated by averaging the entries across trials for the same channel and time position, while the random missing channels were estimated by averaging the corresponding channels across the remaining trials. The AIaT algorithm will be considered the reference in our experiments.

*2.4. Parameter tuning*

Two of the four tensor completion algorithms presented so far have parameters that need to be tuned. The CP-WOPT algorithm requires the tensor rank parameter *R* to be tuned based on available data. On the other side, the 3DPB-TC algorithm, uses dictionaries that must be learned from data signals. Additionally, a sparsity parameter, which is related to the number of atoms of the dictionaries, need to be tuned based on data too (see details in [21]). The other two algorithms, BCPF and HaLRT, are completely parameter-free from the user's point of view because BCPF algorithm automatically tunes all the internal parameters and HaLRT has only one parameter ($rho$) empirically set by the developers to $10^{-7}$. Therefore, this process is transparent to the user.

To select the proper values of all these parameters, we used data from session 4 of the user C (i.e., dataset SubC_s4). We select this session/user because it was the one who produced the best classification rate using all the available points (no missing data, see right-most column in **Table 2**). Using this subject dataset, the best rank R = 250 and sparsity value ρ = 0.1 were found for CP-WOPT and 3DPB-TC algorithms, respectively (see Results, **Figure 4(a-b)**. Once the parameters were tuned the algorithms were tested on the other subject datasets.

## 2.5. *Simulation of corrupted data (mask generation)*

The goal of this work is to evaluate EEG completion as a strategy for dealing with missing or corrupted samples in a BCI experiment. Therefore, we will use a mask to artificially generate missing data in our EEG recordings. This mask will have two different structures. First, we will compare the performance of completion algorithms by using a random entries mask with no predefined structure (random entries case, **Figure 3(a)**). This scenario was already considered in a recent conference contribution [43], where only the BCPF algorithm was used. To evaluate a more realistic scenario for a BCI application, we will also generate a mask constrained to have missing channels in random trials (random channels case, **Figure 3(b)**). The later situation can occur, for example, when an electrode has an improper impedance value or, for any reason, is completely detached during the experiment.

<<FIGURE 3 HERE>>

## 2.6. *BCI Classifier*

As described in the previous section, we use tensor completion algorithms for a BCI experiment, applying masks that simulate several scenarios of missing samples in different trials. To measure the improvement achieved by each algorithm, we implemented Linear Discriminant Analysis (LDA) and linear Support Vector Machine (SVM) classification methods [44] to infer the imagined action performed by the subject based on four features vectors (2 first and 2 last filters) extracted with the classical Common Spatial Pattern (CSP) filtering technique.

CSP filters are designed to find spatial filters so that the variance of the filtered signal is maximum for one class and minimum for the other [45]. Because the variance of a pass-band filtered signal in a specific band is actually the power of this signal in this band, the CSP finds spatial filters that lead to optimally discriminating band power characteristics, since their values would be the maximum of different between classes [46] [45].

These classifiers were used on the original; the average interpolation and the reconstructed data tensor obtained by each of the proposed tensor completion algorithms.

## 2.7. Performance Measures

We evaluate and compare the correctness of the proposed tensor completion algorithms by computing: (*i*) the error of the reconstructed EEG tensor data, and (*ii*) the task classification performance (percentage of correct BCI outputs). Below, we define these measures

### 2.7.1. Normalized Root Mean Square Error (*NRMSE*)

*The Root Mean Square Error* (*RMSE*) is a common measure of the differences between an estimator $\hat{x}$ and the correct values $x$ (EEG signals in our case). Normalizing *RMSE* facilitates the comparison between datasets or models with different scales. The *Normalized RMSE* (*NRMSE*) is defined as follows:

$$NRMSE = \sqrt{\frac{\langle(\hat{x}-x)^2\rangle}{\langle x^2 \rangle}}, \qquad (14)$$

where $\langle \cdot \rangle$ is the expectation operator. The *NRMSE* is always nonnegative and the best possible score is 0. It is also noted its relationship with the Signal to Noise Ratio, i.e., $SNR_{dB} = -20 log_{10}(NRMSE)$. In our experiments, we calculated *NRMSE* values based on all the entries of the tensor data.

### 2.7.2. Classification accuracy evaluation

To ensure statistically stable results of the classification performance, we applied *K*-fold cross-validation. More specifically, we randomly divided the available trials in $K = 10$ groups of trials for each session and each subject. We then trained the classifiers (LDA or SVM, respectively) using $K - 1 = 9$ groups and tested it on the remaining group. We repeated this procedure $N = 100$ times (tests) for different random partitions of data and computed the mean and standard deviation of correct classification trials.

## 3. Results

To demonstrate the usefulness of the proposed EEG data completion methods in a BCI scenario, we did experiments with real EEG tensor datasets, enforcing them to have intentionally corrupted (missing) entries in the following two situations: (1) random missing entries, and (2) random missing channels, as explained in **Figure 3**.

In all the cases, we conducted the experiments using different amounts of missing samples. More precisely, we considered 1%, 5%, 10%, 15% and 20% of missing samples on the whole tensor. We did not consider larger amounts of missing samples because in real applications using six channels, it is not likely to have more than one corrupted channel (1/6 = 16.7%).

In the following section, the results of the parameter tuning based on a single subject are presented, followed by the results obtained with CP-WOPT, 3DPB-TC, BCPF, HaLRT and AIaT algorithms when they are applied to all available subjects not used for parameter tuning (five subjects and seven sessions, see **Table 2**).

### 3.1. Parameter tuning results

To optimally tune the parameters of the CP-WOPT and 3DPB-TC algorithms, we randomly removed 10% of the channels of the whole EEG tensor, as illustrated in **Figure 3(b)**, and performed a grid search over the parameters values using the dataset from session "SubC_s4". More specifically, we computed the NRMSE (equation **(14)**) for a range of parameters values and set the parameter such that the error is minimized. In **Figure 4(a)** and **(b)**, the reconstruction error (*NRMSE*) versus rank $R$ (CP-WOPT) and sparsity $\rho$ (3DPB-TC) are shown, respectively. Additionally, algorithm 3DPB-TC requires to learn a Kronecker dictionary on which patches has a sparse representation. To do, so we applied the dictionary learning algorithm described in [21] to all the 3D patches obtained from the dataset of session "SubC_s4".

### 3.2. Experiments with corrupted/missing data

For all the subjects and sessions, we computed the normalized *RMSE* (*NRMSE*) and the BCI classification performance using the reconstructed tensor obtained with all the algorithms with missing samples ranging from 1% to 20%. The obtained classification performance was compared to the one obtained with the original (complete) EEG data tensor.

First, we made experiments by generating random missing values along all the channels in the EEG tensor without any restriction about the structure of the missed values (**Figure 3(a)**). Even if this is not a realistic scenario, our results demonstrate that the average interpolation and all tensor completion algorithms are able to recover missing entries with a high degree of accuracy, which are consistent with previously reported results [43]. We also performed experiments for a more realistic scenario where complete channels are corrupted/missing. Our results demonstrate that the proposed tensor completion algorithms are still useful in this more difficult scenario.

We report the *NRMSE* values obtained with CP-WOPT (yellow), 3DPB-TC (blue), BCPF (red), and HaLRT (purple) for random missing entries (**Figure 4(c)**) and channels (**Figure 4(d)**). As a reference, we also show the obtained *NRMSE* values using AIaT algorithm (black), i.e., the case where missing entries where replaced by the average interpolation across channels as the simplest way of use the redundancy provided by the trials. It is noted that, in general, all the algorithms helped to considerably reduce the error with respect to the interpolation case. For example, in the case of having 10% of missing entries, the CP-WOPT algorithm reduces the *NRMSE* from 0.308 to 0.056 (82% reduction factor) and from 0.308 to 0.135 (56% reduction factor) for the random missing entries and random missing channels cases, respectively. From these results, we conclude that the case of having random missing channels is more difficult, which gives us less error reduction. It is highlighted that only the BCPF algorithm failed to reduce the error for the case with 20% missing data (missing channels). We think that this effect is because the BCPF assumes a Gaussian model for the model error, which is not accurate for the case of EEG data causing that the algorithm provides a biased estimation of the rank. However, as we will show below, this slightly increase on the reconstruction error (*NRMSE*) does not imply a decrease in classification performance.

In **Figure 4(e)** and **(f)**, we report the classification performance obtained after completing the tensor data for random missing entries and channels, respectively, using the LDA classifier, while in **Figure 4(g)** and **(h)** we report the same results using the SVM classifier. As reference, we also plot the classification performance obtained using the AIaT algorithm (black) and the original data (no missing entries, green). Compared to the AIaT case, all tensor completion algorithms succeed to increase the classification performance. It is highlighted that, even though the completion is more difficult in the missing channels case (**Figure 4(d)**), the improvement of the tensor completion algorithms in the classification performance compared to the AIaT algorithm is better in the missing channels case compared to the missing entries case. It is noted that algorithms BCPF (red) and CP-WOPT (yellow) over-performed 3DPB-TC (blue) in terms of attained classification performance, while HaLRT (purple) obtained the worsts results.

*<<FIGURE 4 HERE>>*

## 4. Discussion and conclusion

In this paper, we have investigated the use of tensor completion algorithms for EEG tensor data in a BCI scenario. Unlike most of the literature on tensor completion, we have applied the algorithms to the realistic case in which some selective EEG channels are corrupted or even missing EEG. This can occur, for example, if an electrode is detached, improperly placed on the skull, or capturing some noise. Traditionally, noisy recordings are discarded and new ones are registered. This can cause delays on the protocols, especially in the calibration step of any BCI procedure, and worst, it generates fatigue and discomfort feeling in the user. In order to avoid such situations, we investigated a way to deal with these noisy channels by completing the data using tensor completion algorithms. This approach is original and different from other strategies like trying to optimize the channels for each user [47]; applying some denoising/artefact removal algorithms [48] [49], or using much more complex classification systems as recently proposed [50].

As showed in **Figure 2(c)**, the once the corrupted/missing data is identified, a mask is applied to the EEG tensor in order to later estimate the corrupted/missing parts. We do not focus in how to decide which samples are corrupted and which are not. This is not a trivial task and has been analysed in different papers. For example, in [51] a method is proposed based on statistical parameters estimated for various aspects of data in both the EEG time series and in the independent components of the EEG. In [17] the authors propose an approach based on the time–frequency shape of muscular artifacts, to achieve a reliable and automatic scoring of them. A method based on clustering is proposed in [52]. The method is based on separation of the ERP recordings into independent components, then to cluster them together with ocular reference signals and finally removing the components in the cluster that contains the ocular reference signals. A semi-automatic artifact rejection system is proposed in [49] in order to detect and reject artifacts. In this case, the identification of artifacts is based on the kurtosis, sample entropy, zero-crossing rate and fractal dimension measures of the sources obtained after applying a blind source separation algorithm on the data. Finally, a fully automated and online artifact removal method is proposed in [53], based on a combination of wavelet decomposition, independent component analysis, and thresholding. We refer the reader to [54] as a recently published reviewer paper regarding requirement for artifact removal in BCI.

Random missing entries (**Figure 3(a)**) are relatively easily retrieved by any of the analysed algorithms, allowing for a small *NRMSE* and, consequently, better classification performance, comparable to the original data. It is noted that noisy data decreased the classification performance up to almost 6% (LDA case) when the percentage of random missing entries was 20% and all algorithms improved the classification performance up to a level comparable to the case of clean EEG data (original EEG tensor) (**Figure 4(e)**). As missing entries are random, the neighbourhood points contain enough information to estimate them, which allows to recover the classification performance. These results are also comparable to the ones obtained in [43], but this scenario is not realistic from the application point of view.

To provide a more realistic scenario, we then investigated the effect of having missing channels (**Figure 3(b)**). This is a more challenging case where all the data points of some channels are missed. When analysing classification performance for random missing channels, results differ. The classification performance based on noisy data decreases at a faster rate. However, all tensor completion algorithms are capable to improve the classification performance with respect to the interpolation case. It is highlighted that interpolated data decreased the classification performance (LDA case) up to almost 20% when the percentage of random missing channels was 10% and algorithms CP-WOPT and BCPF improved the classification performance increasing it by 12%, approximately (**Figure 4(f)**). At 20% of missing channels the results are still over 70% of classification performance for most of the TC algorithms, while the AIaT algorithm (interpolated data) obtains less than 60%. Similar numbers are obtained for the SVM classifier. We note that, regarding the classification performance, BCPF and CP-WOPT outperforms 3DPB-TC and HaLRT, being this last algorithm the worst of all of them. We think that the behaviour of 3DPB-TC algorithm is due because it uses local information (small patches) to reconstruct missing entries while BCPF and CP-WOPT uses a global model for the EEG tensor data.

To our best knowledge this is the first work applying tensor completion algorithms to the realistic scenario of missing channels in BCI, but other applications could use the same strategy. This is the case, for example, of Neuro-Feedback (NF) [55], classification/estimation of emotions [56] or biometric identification [57], in which again we cannot discard frames. We experimentally demonstrated that it is possible to deal with that challenging situation, and instead of discarding the noisy data we can take advantage of it and avoid additional EEG data recordings. On the other hand, we also identified the differences in performance between algorithms depending on what we measure. We observed that the best algorithm for tensor completion, in terms of *NRMSE*, is not the best in terms of obtained classification performance. This fact, surprising at first glance, suggest new lines of research by developing new completion algorithms that maximize the classification performance directly.

**Acknowledgements**

JSC was partially supported by the Spanish Ministerio de Ciencias e Innovación grant TEC2016-77791-C4-2-R and the University of Vic−Central University of Catalonia grant R0947. CFC was supported by NSF IIS-1636893, NSF BCS-1734853, NIH NIMH ULTTR001108 and partially supported by the Indiana University Areas of Emergent Research initiative "Learning: Brains, Machines, Children". QZ was supported by JSPS KAKENHI grant No. 17K00326, NSFC China grant No. 61773129 and JST CREST grant No. JPMJCR1784. AC was partially supported by the MES RF grant 14.756.31.0001 and the Polish National Science Center grant 2016/20/W/N24/00354.

**Compliance with Ethical Standards**

Conflict of Interest: All authors declare that they have no conflict of interest.

Informed Consent: Informed consent was obtained from all participants.

Human and Animal Rights: All experiments were performed in accordance with the tenets of the Declaration of Helsinki.

# TABLES

**Table 1:** Control signals commonly used in EEG BCI systems.
In this paper, we use Motor Imagery control signals.

| Control signal | Physiological phenomena |
|---|---|
| Visual Evoked Potentials (VEP) | Brain signal modulations in the visual cortex |
| Slow Cortical Potentials (SCP) | Slow voltages shift in the brain signals |
| P300 Evoked Potentials | Positive peaks due to infrequent and relevant stimulus |
| Motor Imagery | Sensorimotor rhythms modulations synchronized to motor imagery activities |

**Table 2:** The EEG BCI dataset: a total of eight recording sessions from five subjects (A, C, F, G and H) were collected at LABSP-RIKEN Laboratory using six channels g.tec EEG system. In the rightmost column, the mean classification accuracy and standard deviation (std) over 100 $K$-fold cross validation ($K = 10$) tests, for each of the available 8 datasets, is shown. We used an LDA classifier applied to the original BCI data (no missing values) in all the tests for each dataset. The session with highest accuracy, highlighted in bold, is used for tuning parameters of the algorithms.

| Dataset | Subject | Duration time | Number of trials | Class. Accuracy mean±std |
|---|---|---|---|---|
| SubA_s1 | A | 3s | 130 | 0.88±0.01 |
| SubC_s1 | C | 3s | 170 | 0.86±0.01 |
| SubC_s2 | C | 3s | 158 | 0.85±0.01 |
| SubC_s3 | C | 3s | 120 | 0.89±0.01 |
| **SubC_s4** | **C** | **3s** | **90** | **0.93±0.01** |
| SubF_s1 | F | 4s | 80 | 0.71±0.03 |
| SubG_s1 | G | 4s | 120 | 0.81±0.01 |
| SubH_s1 | H | 3s | 150 | 0.71±0.02 |

**Table 3:** List of algorithms used in the experiments.

| Name | Acronym | Parameters to be determined |
|---|---|---|
| CP Weighted Optimization [22] | CP-WOPT | Tensor rank $R$ |
| 3D Patch-based Tensor Completion [21] | 3DPB-TC | Dictionary and sparsity parameter $\rho$. |
| Bayesian CP factorization for tensor completion [24] | BCPF | None |
| High accuracy Low Rank Tensor Completion [25] | HaLRT | $\rho$ (empirically $\rho = 10^{-7}$) |
| Average interpolation across trials | AIaT | None |

**FIGURES**

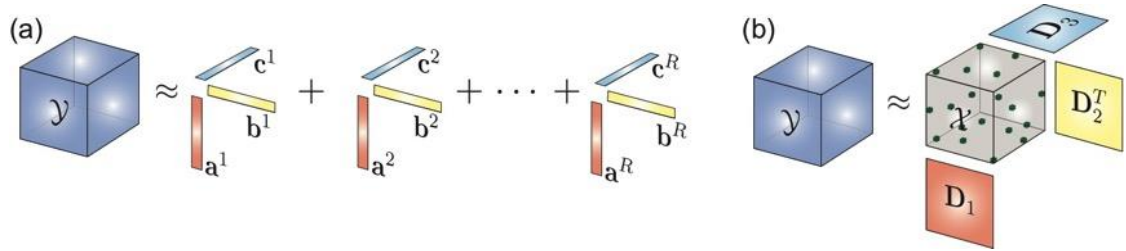

**Figure 1:** Tensor decompositions: (a) CANonical DECOMPosition (CANDECOM). (b) Sparse Tucker model

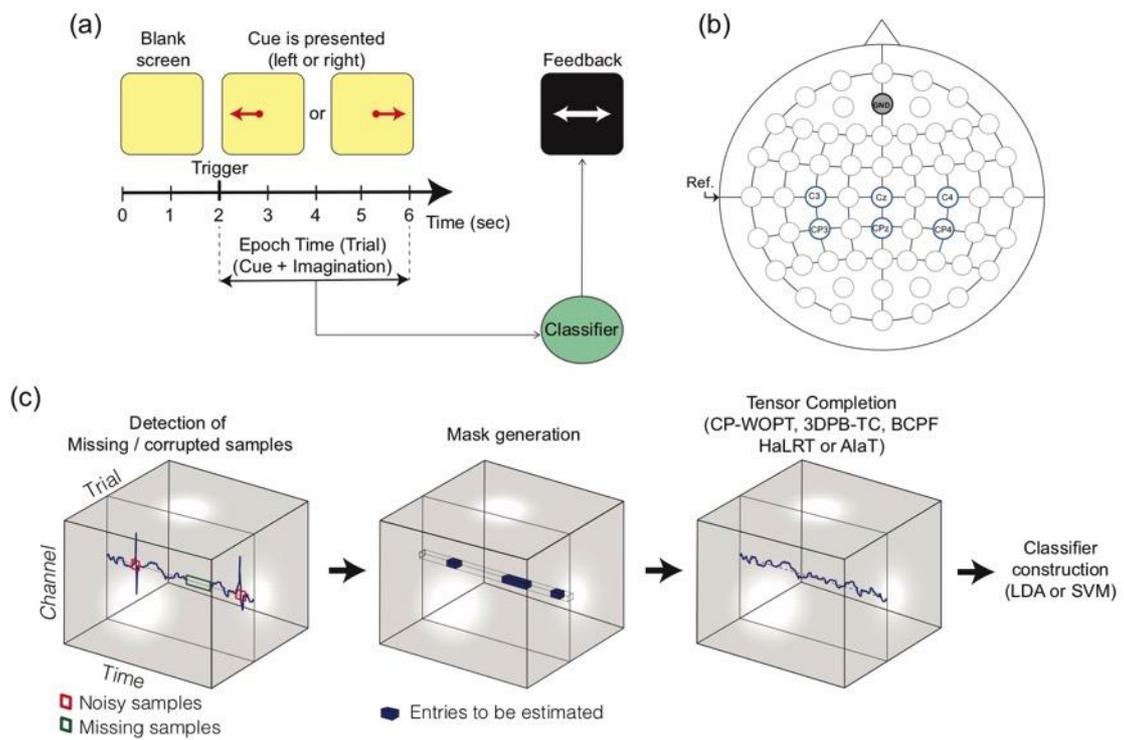

**Figure 2:** BCI experiment description: (a) Timing scheme for the MI paradigm. (b) Electrode locations. (c) Sequence of operations performed before calibration.

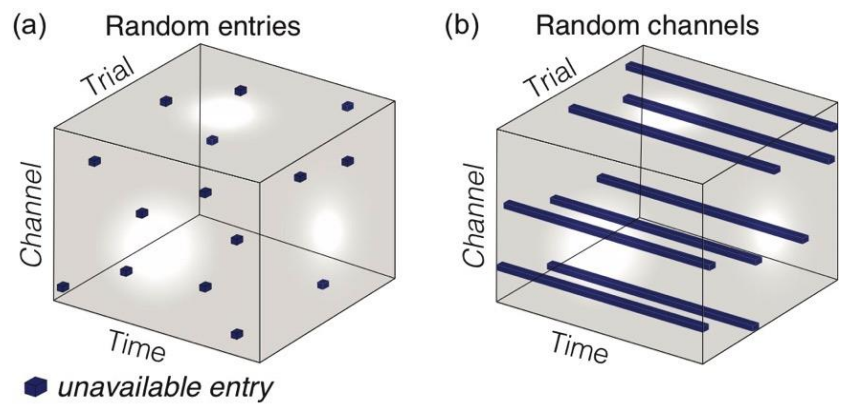

**Figure 3:** Examples of masks used for generating observed EEG data (with random missing entries). (a) Example of uniformly distributed random mask generated without any constrains. (b) Example of random mask generated with random missing channels.

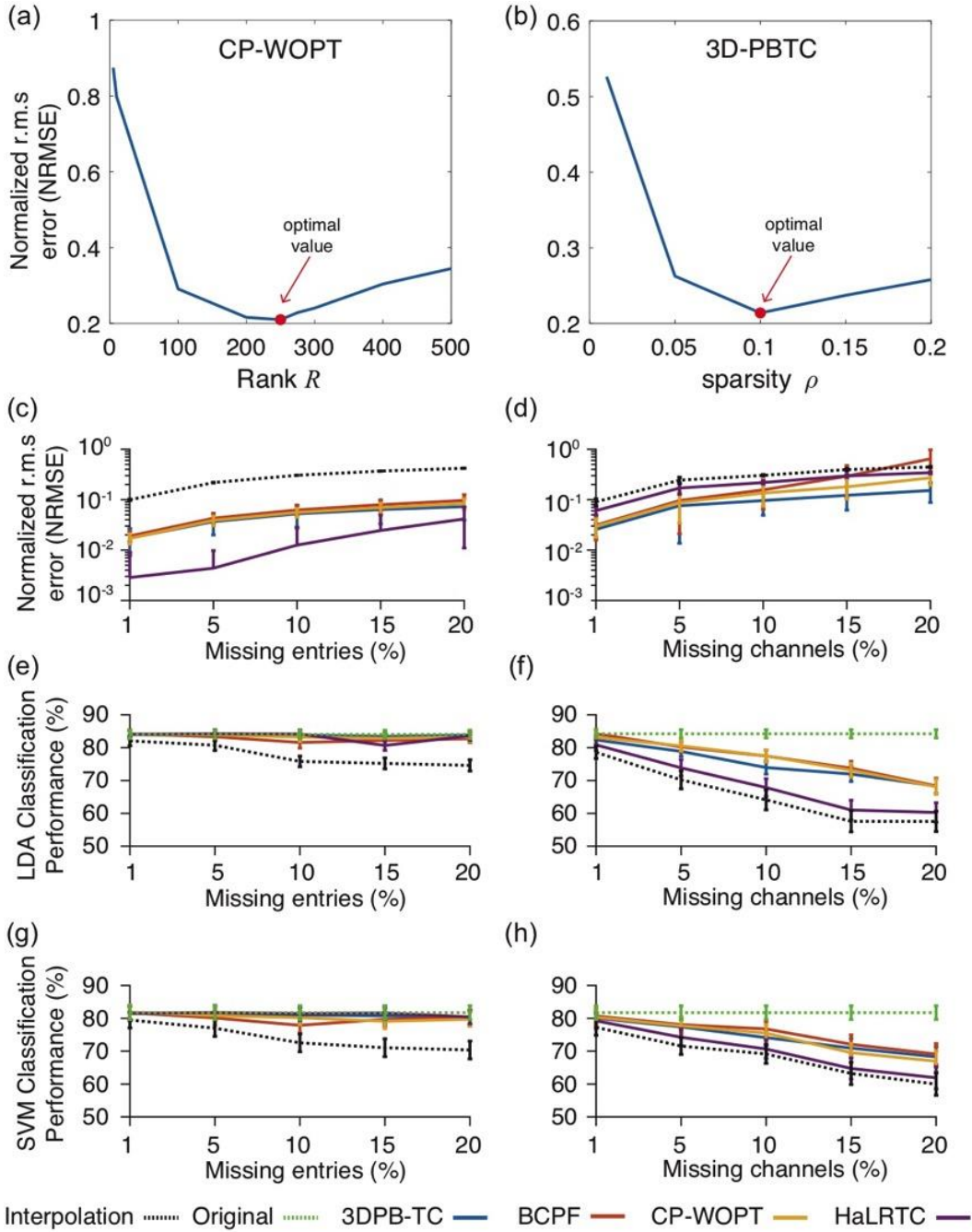

**Figure 4:** Experimental results: (a) NRMSE evolution as a function of Rank $R$ in CP-WOPT algorithm. Best result was obtained for $R$=250. (b) NRMSE evolution as a function of sparsity $\rho$ for 3DPBTC algorithm. Best result was obtained for $\rho$=0.1. (c) NRMSE evolution as a function of percentage of missing entries for the case of random missing entries. (d) NRMSE evolution as a function of percentage of missing entries for the case of random missing channels. (e) LDA classification performance (%) evolution as a function of percentage of missing entries for the case of random missing entries. (f) LDA classification performance (%) evolution as a function of percentage of missing entries for the case of random missing channels. (g) and (h) Same as (e-f) for SVM classifier.